\documentclass[pra,twocolumn,showpacs]{revtex4}
\input epsf

\begin{document}

\title{Peculiar Features of Entangled States with Post-Selection}

\author{Yakir Aharonov$^{1,2}$}
\author{Shmuel Nussinov$^{1,2}$}
\author{Sandu Popescu$^{3}$}
\author{Lev Vaidman$^{1,2}$}
\affiliation{$^1$ School of Physics and Astronomy, Tel Aviv University,  %
Tel Aviv, Israel} \affiliation{$^2$ Institute for Quantum Studies, Chapman University, 1 University Dr, Orange, CA, USA}
\affiliation{$^3$ H.H.Wills Physics Laboratory, University of Bristol, %
 Tyndall Avenue, Bristol BS8 1TL, U.K.}

\date{\today}

\begin{abstract}
We consider  quantum systems in entangled states
 post-selected in non-entangled states. Such systems exhibit
unusual behavior, in particular when weak measurements are performed at
 intermediate times.
\end{abstract}
\pacs{03.65.Ta}

\maketitle
Measurements performed on pre- and post-selected
quantum systems often exhibit peculiar results. One particular example is that of a single particle
which is found with certainty in any one of a large number of
boxes if only this box is opened \cite{AV91,AV03}. In spite of experimental implementations of these ideas \cite{RLS,Kolend,George}, there is still some controversy about this example
\cite{SS,DTSQT,Grif,Kent,Kast,reply,SHI,CO,CO-co,Kast2,Kirk,Lei,RV,Kirk3,Nativ}. Even more surprising is the fact that
outcomes of
weak measurements \cite{AV90}, namely, standard von Neumann measurements
 with weakened interaction  yield {\it weak values}
which might be {\it far away} from the range of possible eigenvalues. This feature led to practical applications for high precision measurements \cite{HK,howel}. Here
we  consider the peculiar features of quantum systems
when the  pre-selected state is entangled.

First, let us consider the case when the pre-selected state is the following Greeberger-Horne-Zeilinger (GHZ)-like entangled state of $N$
spin-${1\over 2}$ particles:
\begin{equation}
  \label{eq:ghz}
|\Psi \rangle_1 = {1\over \sqrt 2}\big (\prod_{n=1}^N  |{\uparrow}_z
\rangle_n + \prod_{n=1}^N  |{\downarrow}_z \rangle_n \big ).
\end{equation}
Just like the original three-particle GHZ state, it can be used to show that no local hidden variable
theory is consistent with the predictions of quantum theory.

 To this end consider the following sets of measurements. The first set consists of $\sigma_x$ measurements performed over all particles. The second set consists of $\sigma_x$ measurements
for all  particles except for two particles for which $\sigma_y$   are measured. Since
\begin{equation}
  \label{eq:eigen}
\prod_{n=1}^N \sigma_x^{(n)}  |\Psi \rangle_1 = |\Psi \rangle_1 ,
\end{equation}
and
\begin{equation}
  \label{eq:eigen2}
\prod_{n\neq k,l}^N \sigma_x^{(n)} \sigma_y^{(k)} \sigma_y^{(l)}  \ |\Psi
\rangle_1 = - |\Psi \rangle_1 ,
\end{equation}
the outcomes  of the two sets of measurements should fulfill:
\begin{equation}
  \label{eq:eq1}
\prod_{n=1}^N \sigma_x^{(n)}=1,
\end{equation}
and
\begin{equation}
 \label{eq:eq2}\prod_{n\neq k,l}^N
\sigma_x^{(n)} \sigma_y^{(k)} \sigma_y^{(l)} =-1 ,
\end{equation}
respectively.

Next consider three particular cases of equation (5) where the pairs of particles for which $\sigma_y$ was measured were chosen from a triplet of   particles $s, t$, and $r$.

We then have
\begin{eqnarray}
  \label{eq:4eq}
\prod_{n=1}^N \sigma_x^{(n)}=1, \nonumber\\
 \prod_{n\neq s,t}^N \sigma_x^{(n)} \sigma_y^{(s)} \sigma_y^{(t)} =-1,  \nonumber\\ \prod_{n\neq s,r}^N \sigma_x^{(n)} \sigma_y^{(s)} \sigma_y^{(r)} =-1,\\ \prod_{n\neq t,r}^N \sigma_x^{(n)} \sigma_y^{(t)} \sigma_y^{(r)} =-1.\nonumber
\end{eqnarray}
Had there been local hidden variables, then all variables should have
values, i.e., prior to the measurement every particle should ``know''
the value of all spin components and, in particular, the $x$ and the
$y$ components. These values should fulfill all four equations
(\ref{eq:4eq}).  Just like in the original GHZ case, this is impossible since the product of the
left hand sides is the product of squares (or fourth powers) of the spin
variables and the product of the right hand sides is $-1$.

In order to see surprising results we will consider the particular
post-selected state $|\Psi \rangle_2$ in which all spins were found with $\sigma_x =1$,
i.e.,
\begin{equation}
|\Psi \rangle_2 =\prod_{n=1}^N |{\uparrow}_x\rangle_n.
\label{post1}
\end{equation}
A unique feature of this particular pre- and post-selection is
that at  intermediate times no pair of particles can be found with the
same $\sigma_y$, otherwise, the requirement (\ref{eq:eq2})
cannot be fulfilled.

To see this in a more dramatic way we map  states of $N$ spin-${1\over
  2}$ particles into the states of $N$ distinguishable particles
which can reside in two spatially separated boxes $A$ and $B$:
 \begin{equation}
  \label{eq:tran}
|{\uparrow_y}\rangle \equiv  |A \rangle, ~~|{\downarrow}_y\rangle \equiv  |B \rangle,
\end{equation}
where $|A \rangle$  ($|B \rangle$) is the state of the particle in box
$A$ ($B$).
Thus, we have $N$ (which might be as large as we want) particles in two
boxes, but if we try to find any specific pair of particles in one
box we are bound  to fail! This is a very paradoxical situation for
$N\geq 3$. The probability of the post-selection which leads to this
situation is $1\over 2^N$, but for $N$ which is not too large it
is  feasible to implement experimentally.

The ``paradox'' in this example follows from i) the obvious classical observation that
when three or more particles are located in two boxes, then at least one pair
has to be in one box and ii) the quantum property which follows from the joint
pre-and post-selection (but not from pre- or post-selection
separately) that no pair of particles can be found in a single
box.

 Hardy has already proposed \cite{Hardy} a related experiment on a
pre- and post-selected system. It can be
viewed  as an experiment with two particles in two boxes \cite{A-Ha}. The particles
were prepared in the entangled state
\begin{equation}
  \label{eq:ABC}
|\Psi_1\rangle =  {1\over \sqrt 3}(|A \rangle_1 |B\rangle_2  +|B \rangle_1 |A\rangle_2 + |B \rangle_1 |B\rangle_2).
\end{equation}
The preparation started with the non-entangled state
\begin{equation}
  \label{eq:ABCnon}
|\Psi\rangle =  {1\over  2}(|A \rangle_1 +|B\rangle_1)\ ( |A \rangle_2 + |B\rangle_2),
\end{equation}
and entanglement was achieved by projecting out the state $|A \rangle_1
|A\rangle_2$. Then, it was post-selected in the state
\begin{equation}
\label{eq:ABCpost}
|\Psi_2\rangle =  {1\over  2}(|A \rangle_1 -|B\rangle_1)\ ( |A \rangle_2 - |B\rangle_2).
\end{equation}

The paradoxical feature here is that at intermediate times each particle
is  found in box $A$ if it is searched for
there, but the two particles cannot be found together in $A$. The fact that each
particle is (in this particular sense) in $A$ follows from the joint pre-
and post-selection, while the fact that both are not there, follows
directly from the pre-selection.

A generalization of this  for the case of $N$ particles is the
pre-selection of the state
\begin{equation}
  \label{eq:ABN}
|\Psi_1\rangle  =  {1\over \sqrt {N^2-N+1}} \big [ (N-1)\prod_{n=1}^N  |B\rangle_n
+\sum_{n=1}^N |A \rangle_n \prod_{j\neq n}^N  |B\rangle_j \big ],
\end{equation}
and the post- selection of the state
\begin{equation}
\label{eq:ABCpostN}
|\Psi_2\rangle =  {1\over \sqrt {2^N}}\prod_{n=1}^N  (|A \rangle_n -|B\rangle_n).
\end{equation}
Every one of the $N$ particles is  found with unit probability in box $A$ if it is searched for
there. However,  not only that they  cannot all be found in $A$,  any
number of particles larger than one cannot be found there.

All these peculiarities of pre- and
post-selected systems can be seen, albeit in a much more complicated way, in the standard
formalism of quantum mechanics with a single quantum state evolving
from the past to the future. The most surprising features
of pre- and post-selected quantum systems are manifested in the
outcomes of {\it weak measurements} \cite{AV90}. Any standard measuring
procedure with weak enough coupling performed on a quantum system
pre-selected in state $|\Psi\rangle_1$  and post-selected in state
$|\Psi\rangle_2$ yields, instead of one of the eigen values of the
measured observable $O$, the {\it
  weak value} of $O$ given by the following expression:
\begin{equation}
O_w \equiv { \langle{\Psi_2} \vert O \vert\Psi_1\rangle
\over \langle{\Psi_2}\vert{\Psi_1}\rangle } .
\label{wv}
\end{equation}
The weak value might be far away from the eigenvalues of
$O$ and the standard quantum formalism can explain the outcome by
surprising universal interference phenomenon of the pointer of the
measuring device.

The following two theorems connect weak and ``strong'' (standard von Neumann) measurements
 \cite{AV91}: i) if, for a pre- and
post-selected system the outcome $o_i$ of the
strong measurement of $O$ is known with certainty, then the weak value
is the same $O_w =o_i$; ii)   if, for a pre- and
post-selected system the weak value of a dichotomic variable $O$ is equal
to one of the eigenvalues $o_i$, then a strong measurement of $O$ will
yield this value $o_i$ with certainty.

In the above generalized examples with $N$ particles,  weak measurements suggest that, in some sense, all
particles are in box $A$ in spite of the fact that at most one
particle can be found in $A$. Indeed, the additive property of weak values implies:
\begin{equation}
\big ( \sum_{n=1}^N {\rm \bf P_A^{({\it n})}}\big )_w  =  \sum_{n=1}^N \big ({\rm
  \bf P_A^{({\it n})}}\big )_w = N.
\end{equation}
Note, however that the weak measurement of the product of the
projection operators vanishes,
\begin{equation}
\big ( \prod_{n=1}^N {\rm \bf
  P_A^{({\it n})}}\big )_w =0 ~\big [\neq \prod_{n=1}^N \big ( {\rm \bf
  P_A^{({\it n})}}\big )_w~\big ],
\label{produ}
\end{equation}
 because the strong value of the product vanishes with certainty.

Let us return to the example with $N$ pre- and post-selected particles
in two boxes in which every particular pair of particles cannot be
found in the same box. It is instructive to consider this example for identical spinless bosons. The statement that strong
measurement cannot find any particular
pair of particles in any  single box  is
then meaningless because there is no  ``particular pair'' among $N$ identical particles. Nevertheless, we still can make a statement
about weak measurements. Let us assume that when two particles are
in the same box, there is a potential $V$
between them.  Weak measurement of the interaction energy between
$N$ particles in the two boxes in our pre- and post-selected state should then
yield zero! This can be seen in a straightforward way by calculating
the weak value of the potential energy $\sum_{n\neq m} V ~{\rm \bf
  P_A^{({\it n})} \bf P_B^{({\it m})}}$ for a system pre-selected in  state (\ref{eq:ghz})
and post-selected in state (\ref{post1}).  According to the mapping  (\ref{eq:tran})  the particles are described (up to
normalization)  by the following two-state vector
\begin{equation}
  \label{tsv1}
 \prod_{n=1}^N (\langle A|_n + \langle B|_n)~~\big
 (\prod_{n=1}^N ( |A \rangle_n - i|B\rangle_n) + \prod_{n=1}^N ( |B
 \rangle_n - i|A\rangle_n)\big ).
\end{equation}

Some intuition and an alternative  proof of this
result is as follows. The potential
energy is the same for identical and nonidentical particles. For
nonidentical particles, the measurement of the potential energy of any
specific  pair  yields zero for this particular pre- and
post-selection. Therefore, the weak value of the potential energy for
this pair is zero. Weak values are additive: $(A +B)_w= A_w
+B_w$. Thus, the weak value of  the potential energy of all pairs
together is zero. Therefore, the weak value  of the potential energy
of $N$ identical particles  is also zero.

Our next example is based on the most famous entangled state, the Einstein-Podolsky-Rosen-Bohm
state (EPR), which is a singlet state of two spin-$1\over 2$ particles. It
has been noted \cite{V93} that this pre-selected state when post-selected with
a particular  product state
has unusual property at intermediate times. The two state vector is:
\begin{equation}
\langle{\uparrow}_x|_1\langle{\uparrow}_z|_2~~~ {1\over \sqrt 2}( |{\uparrow}_z \rangle_1~|{\downarrow}_z \rangle_2 -|{\downarrow}_z \rangle_1~|{\uparrow}_z \rangle_2 ) .
\label{tsvxz}
\end{equation}
 We know that the following outcomes are obtained with certainty if only one of the measurements is performed:
$\sigma_z^{(1)} =- 1$,  $\sigma_x^{(2)}= -1$, and  $\sigma_z^{(1)}
\sigma_x^{(2)}= -1$. This demonstrates the failure of the ``product rule''
for pre- and post-selected quantum systems. If we consider  the
spatial dichotomic variable instead of the spin with the correspondence:
 \begin{eqnarray}
  \label{eq:tran2}
|{\uparrow_z}\rangle_1 \equiv  |A \rangle_1, ~~|{\downarrow}_z\rangle_1
\equiv  |B \rangle_1,\\
|{\uparrow_x}\rangle_2 \equiv  |A \rangle_2, ~~|{\downarrow}_x\rangle_2 \equiv  |B \rangle_2,
\end{eqnarray}
then again we reach the situation in which each particle is with
certainty to be found in box $B$ if it is searched for there, but they
cannot be found in $B$ together.

We can see another amusing feature of the two-state vector (\ref{tsvxz}) when we recast it considering just one spin-$1\over 2$
particle. In this case we map the spin variable of the first EPR
particle into  the spatial variable of our particle via
(\ref{eq:tran2}) and the spin of the second EPR particle becomes the spin of our particle. With this correspondence, the particle is pre-selected
in the state
\begin{equation}
|\Psi_1 \rangle = {1\over \sqrt 2}( |A \rangle |{\downarrow}_z\rangle
  - |B \rangle |{\uparrow}_z \rangle),
\label{EPR1p}
\end{equation}
and post-selected in the state
\begin{equation}
|\Psi_2 \rangle  = {1\over \sqrt 2}( |A \rangle + |B \rangle )~
|{\uparrow}_z \rangle .
\label{post}
\end{equation}
In this pre- and post-selected state the probability to find the particle in $A$ at
intermediate times  vanishes. The weak value $({\bf \rm P_A})_w$ vanishes so that, in some sense, the particle is not in
$A$. Nevertheless, the weak measurement of the spin component $\sigma_x$ in $A$
yields $({\rm \bf P_A}\sigma_x)_w =-1$. If the particle is an electron or a neutron, we
will then sense a non-vanishing magnetic field in $A$ (if it is weakly
measured), but the weak value of the number of particles in $A$ is
zero. Thus, for ``weak measurement reality'' \cite{WER}  the particle is in $B$,
but its magnetic field is in $A$ which might be arbitrary far from
$B$. For a possible interpretation of this result see \cite{Cat}.

 In the above case, for weak coupling the particle is in $B$, but its
 magnetic field is in $A$. However, the particle generates magnetic field in $B$
 too. By  considering  a particle in a {\it generalized two-state vector} \cite{AV91},
 we  can arrange that the magnetic field  appears only in $A$.
 A particle is described by a generalized  two-state vector $\sum_i \alpha_i \langle
 \Phi_i |~~|\Psi_i \rangle$ when it and an ancilla (which nobody touches
 at the intermediate time) are pre- and post-selected in particular entangled states. The weak value of an observable $O$ of the particle in this case is:
\begin{equation}
  \label{wv-gen}
 O_w = {{\sum_i \alpha_i \langle \Phi_i | O | \Psi_i \rangle }
\over{\sum_i \alpha_i \langle \Phi_i  | \Psi_i \rangle }} .
\end{equation}
 The generalized two-state vector which leads to the phenomena described above is:
 \begin{eqnarray}
(\langle  A| +\langle B| ) \langle{\uparrow}_z | ~~(|{\downarrow}_z \rangle |A
\rangle
  + |{\uparrow}_z \rangle |B \rangle ) + \nonumber \\
 (\langle  A| +\langle B| )
 \langle{\downarrow}_z |
 ~~( |{\uparrow}_z \rangle |A \rangle  +|{\downarrow}_z \rangle |B \rangle
  ).
\label{gen2sv}
\end{eqnarray}
Indeed, it is straightforward to see that
\begin{equation}
({\rm \bf P_A})_w =0,~~~~
({\rm \bf P_B})_w =1.
\label{sigwe}
\end{equation}
Nevertheles, weak measurement can sense the particle's magnetic field only in $A$:
\begin{equation}
({\rm \bf P_A}\sigma_x)_w = 1,~~
({\rm \bf P_A}\sigma_y)_w =({\rm \bf P_A}\sigma_z)_w = 0,
\label{sigweB}
\end{equation}
\begin{equation}
({\rm \bf P_B}\sigma_x)_w =
({\rm \bf P_B}\sigma_y)_w =({\rm \bf P_B}\sigma_z)_w = 0,
\label{sigwe}
\end{equation}
i.e., there is a magnetic field along the $x$ axis corresponding to
 a particle in $A$ with spin $\sigma_x=1$.

We hope that the examples presented above will help to develop an intuition for understanding  pre- and post-selected systems with entanglement and will lead to useful applications of the peculiar effects such system exhibit.

We thank  Eli Cohen for useful discussions.This work was supported in part by the Binational Science Foundation Grant 32/08 and the Israel
Science Foundation  Grant No. 1125/10.

\end{document}